\documentclass[aps,twocolumn,preprintnumbers,superscriptaddress,nofootinbib,a4paper,longbibliography,twoside,floatfix,altaffilletter,tightenlines,showpacs,notitlepage,nofootinbib]{revtex4-2}

\usepackage{amsmath,amssymb} % Maths
\usepackage{graphicx} % Figures
\usepackage{titlesec}
\usepackage[export]{adjustbox} % Vertical alignment of figures
\usepackage{url} % URL formatting
\usepackage[dvipsnames]{xcolor} % Colors
\usepackage{siunitx} % Proper unit formating
\usepackage{etoolbox,orcidlink} % for ORCIDs behind author names
\usepackage[normalem]{ulem}

\makeatletter

\renewcommand{\p@subsection}{}
\makeatother

\titleformat*{\section}{\centering\bfseries\uppercase}
\titlelabel{\thetitle\quad}
\titleformat*{\paragraph}{\bfseries}
\titlespacing*{\paragraph}{0pt}{3.25ex plus 1ex minus .2ex}{1em}

\usepackage{titlesec}
\usepackage{microtype}
\titlelabel{\thetitle\quad}  % remove dot behind section number

\usepackage{hyperref} % References
\usepackage[capitalize]{cleveref} % Nice referencing
\newcommand\myshade{80} % Reference coloring
\colorlet{mylinkcolor}{ForestGreen}
\colorlet{mycitecolor}{Red}
\colorlet{myurlcolor}{violet}
\hypersetup{
  linkcolor  = mylinkcolor!\myshade!black,
  citecolor  = mycitecolor!\myshade!black,
  urlcolor   = myurlcolor!\myshade!black,
  colorlinks = true
}

\newcommand{\github}[1]{\href{https://github.com/#1}{\includegraphics[width=8pt]{Plots/github.pdf}}}

\begin{document}

%=============================================================================
\title{High-Frequency Gravitational Wave Constraints from Precision Spectroscopy}

\newcommand{\cern}{Theoretical Physics Department, CERN,
                   1211 Geneva 23, Switzerland}
\newcommand{\jgu}{Johannes Gutenberg-Universit\"at Mainz, 55128 Mainz, Germany}
\newcommand{\him}{Helmholtz-Institut Mainz, 55099 Mainz, Germany}

\author{Dmitry Budker \orcidlink{0000-0002-7356-4814}\,}
\email{budker@uni-mainz.de}
\affiliation{\him}
\affiliation{\jgu}
\affiliation{GSI Helmholtzzentrum f\"ur Schwerionenforschung GmbH, 64291 Darmstadt, Germany}
\affiliation{Department of Physics, University of California, Berkeley, California 94720-7300, USA}

\author{Valerie Domcke \orcidlink{0000-0002-7208-4464}\,}
\email{valerie.domcke@cern.ch}
\affiliation{\cern}

\author{Joachim Kopp \orcidlink{0000-0003-0600-4996}\,}
\email{jkopp@cern.ch}
\affiliation{\jgu}

\author{Oleg Tretiak \orcidlink{0000-0002-7356-4814}\,}
\email{oleg.tretiak@uni-mainz.de}
\affiliation{\him}
\affiliation{\jgu}

\date{\today}

\preprint{CERN-TH-2026-168, MITP-26-033} 
%=============================================================================

\begin{abstract}
\noindent
Gravitational waves affect the propagation of electromagnetic waves in laser cavities, modulating the frequency of emitted photons.
We use this effect to search for high-frequency gravitational waves between \SI{100}{kHz} and \SI{100}{MHz} using optical precision spectroscopy. Our limits constrain much of this frequency range for the first time. We discuss future improvements of the technique, which we expect to enhance the sensitivity by eight orders of magnitude, and to extend the frequency coverage up to at least \SI{1}{GHz}.
\end{abstract}

%=============================================================================
\maketitle
%=============================================================================

%=============================================================================
\section{Introduction}
\label{sec:intro}
%=============================================================================

Over the past decade, the advent of gravitational wave (GW) astronomy has transformed not only astrophysics, but also related fields such as cosmology and astroparticle physics, with its impact reaching even into nuclear and quantum physics. Most observations target GW frequencies from kilohertz down to nanohertz, where most known and expected astrophysical sources are found. However, there is a diverse and fast-growing community focusing instead on \emph{high-frequency gravitational waves} (HFGW), that is, GW with frequencies above \SI{1}{kHz}. Potential sources of HFGW include primordial black hole mergers~\cite{Franciolini:2022htd}, ultralight ($\ll \si{eV}$) boson superradiance around (primordial) black holes~\cite{Arvanitaki:2010sy, Brito:2015oca}, as well as cosmological processes in the very early Universe~\cite{Aggarwal:2025noe}. In addition to these exotic sources, also a first-order quantum chromodynamics (QCD) phase transition inside a neutron star could emit gravitational waves at $\mathcal{O}(\si{MHz})$ frequencies~\cite{Blas:2022xco, Bleau:2026ala}.

Numerous experimental techniques are under consideration for detecting HFGWs, see Ref.\,\cite{Aggarwal:2025noe} for a comprehensive review. Concrete limits have so far been obtained from interferometers \cite{Holometer:2016qoh, Martinez:2020cdh}, bulk acoustic wave devices \cite{Goryachev:2014yra}, and axion detectors~\cite{Ejlli:2019bqj, Kim:2025izt, Pappas:2025zld}. Many other detection concepts are under active consideration, most of them exploiting the coupling of gravitational waves to electromagnetism~\cite{Ejlli:2019bqj, Berlin:2021txa,Domcke:2022rgu,Bringmann:2023gba,Domcke:2023bat,Navarro:2023eii,Domcke:2024eti}.
Of significant current interest is also the possibility to combine multiple detectors to a global network \cite{Amaral:2026bef}.

Here, we focus on a particularly straightforward technique: a gravitational wave can modulate the proper distance of a photon roundtrip in a laser cavity.
The resonant boundary condition of the cavity translates this into a frequency modulation of the emitted light.
This frequency shift can be detected using precision spectroscopy \cite{Loeb:2015ffa, Vutha:2015aza, Kolkowitz:2016wyg, Bringmann:2023gba}. Using experimental results previously described in Refs.\,\cite{Tretiak:2022ndx, Budker:2024bzj} in the context of wave-like dark matter searches, we set limits on persistent high-frequency gravitational wave signals across a wide frequency range (\SI{20}{kHz}--\SI{100}{MHz}), much of which has not been experimentally explored before. We also argue that significant improvements both in frequency coverage and in sensitivity will be possible over the coming years. The analytical calculations and data analysis routines used in this paper can be found on GitHub \cite{Budker:2026github}.

%=============================================================================
\section{Gravitational-Wave-Induced Photon Frequency Shifts}
\label{sec:theory}
%=============================================================================

\begin{figure}
    \centering
    \includegraphics[width=\columnwidth]{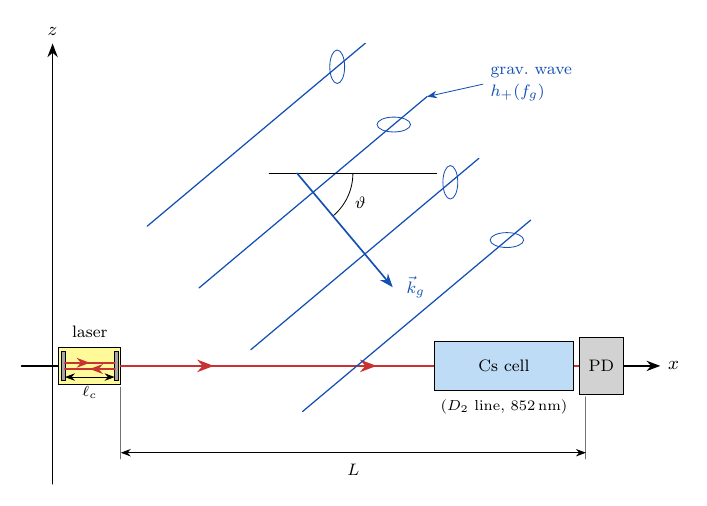}
    \caption{Schematic of a spectroscopic search for high-frequency gravitational waves. A laser with a cavity of size $\ell_c$ produces nearly monochromatic photons tuned to the Cs $D_2$ line at \SI{852}{nm}. The photons propagate along the $x$-axis through a cesium vapor cell to a photodetector (PD), over a baseline $L$. A passing GW (blue parallel lines with the alternating ellipses illustrating the polarisation pattern) affects the laser's resonance frequency according to \cref{eq:fshift-laser}, and further shifts the photon frequency during propagation according to \cref{eq:fshift_freefall}. Note that the vapor cell could be replaced with a different high-quality optical resonator, for instance a Fabry--P\'erot cavity.}
    \label{fig:exp-sketch}
\end{figure}

Our starting point is the experimental setup sketched schematically in \cref{fig:exp-sketch}. Its main components are a laser with a cavity of length $\ell_c$, and a narrow optical resonator a distance $L$ away. In our case, the resonator is a cesium vapor cell, and the laser frequency is tuned to the rising or falling edge of a narrow atomic transition in cesium. The transmissivity of the vapor cell then depends sensitively on the detuning between the laser frequency and the peak frequency of the atomic transition. We assume that photons propagate along the $x$-axis, and we parametrize the gravitational wave as a metric perturbation of the form
\begin{align}
    h^{TT}_{11}(x^\mu) = h_+ s^2_\vartheta \, \cos\left[ 2\pi f_g (t - c_\vartheta x
                                    - s_\vartheta z) + \varphi_0(t) \right] \,,
    \label{eq:htt11}
\end{align}
where $h^{TT}_{11}(x^\mu)$ denotes the first-order correction to the $(xx)$ element of the metric tensor in the transverse--traceless (TT) frame, $h_+$ is the amplitude of the gravitational wave, $f_g$ is its frequency, and $\varphi_0(t)$ its phase at $\vec{x}=0$. We take the wave to propagate in the $x$--$z$ plane at an angle $\vartheta$ relative to the $x$-axis, and we use the shorthand notation $s_\vartheta \equiv \sin\vartheta$, $c_\vartheta = \cos\vartheta$, and $x^\mu \equiv (t, x, y, z)$. Here and in the following, we set $c = \hbar = 1$. Note that, as shown in Ref.\,\cite{Bringmann:2023gba}, the only element of the metric tensor affecting photons propagating along the $x$-direction is $h^{TT}_{11}(x^\mu)$. Notably, the setup in \cref{fig:exp-sketch} is insensitive to cross-polarized ($\times$) gravitational waves. (We use here the convention that the $+$ polarization is defined by one of the two quadrupole axes being aligned with the $y$ axis.)

We consider first the photons inside the laser cavity. Due to the gravitational wave, they experience a phase shift $\Delta\phi|_{\rm laser}$ during one round-trip in the cavity, which can be calculated along the lines of Refs.\,\cite{Maggiore:2007ulw, Domcke:2024abs}. In response to this shift, the laser frequency adjusts in such a way that the appropriate boundary conditions are maintained. Our experimental setups employs a ring laser, in which case the relevant boundary condition is periodicity: the photon phase accumulated during one round trip has to be an integer multiple of $2\pi$. For simplicity, we assume a cavity of length $\ell_c$ oriented along the $x$-axis as shown in \cref{fig:exp-sketch}. Neglecting the short segments of the photon path aligned in the $z$-direction, the frequency shift required to maintain periodic boundary conditions is \cite{Domcke:2024abs}
\begin{multline}
    \hspace{-0.4cm}
    \frac{\Delta f_\gamma}{f_\gamma}\Big|_{\rm laser} \!\!
      = \frac{\Delta\phi|_{\rm laser}}{4\pi \ell_c f_\gamma}
      = \frac{h_+}{8 \pi \ell_c f_g} \Big\{
            (c_\vartheta - 1) \sin[\varphi_0(t) + 4 \pi f_g \ell_c]
                                                \\[0.2cm]
          + (c_\vartheta + 1) \sin[\varphi_0(t)]
          - 2 c_\vartheta \sin[\varphi_0(t) + 2\pi f_g \ell_c (1 - c_\vartheta)]
        \Big\}\,.
    \label{eq:fshift-laser}
\end{multline}
Note that, for the laser to be able to adjust accordingly, the emission line of the gain medium must be broad compared to \cref{eq:fshift-laser}.

In addition, photons also experience a frequency shift during propagation from the laser to the vapor cell \cite{Kaufmann:1970, Estabrook:1975jtn, Tinto:1998ee, Armstrong:2006zz, Lobato:2021ffr, Kolkowitz:2016wyg, Bringmann:2023gba}. This additional frequency shift is given by \cite{Kolkowitz:2016wyg, Bringmann:2023gba}
\begin{multline}
    \hspace{-0.3cm}
    \frac{\Delta f_\gamma}{f_\gamma}\Big|_{\rm prop}
        = \frac{d\Delta\phi|_{\rm prop}}{d(2\pi f_\gamma t)} \\[0.2cm]
        = \frac{h_+}{2} (1 + c_\vartheta)
        \Big\{\!
            \cos[\varphi_0(t)]
          - \cos\big[2\pi f_g L (1 - c_\vartheta) + \varphi_0(t) \big]
        \Big\} \,.
    \label{eq:fshift_freefall}
\end{multline}
Note that the phase shifts underlying \cref{eq:fshift-laser,eq:fshift_freefall} represent the same physical effect as exploited by gravitational wave interferometers such as LIGO \cite{LIGOScientific:2014pky}, Virgo \cite{VIRGO:2014yos}, and KAGRA \cite{KAGRA:2020tym}: a periodic change in the proper distance the photons travel results in a phase modulation (which can equally be interpreted as a frequency modulation~\cite{Domcke:2024abs}). However, the consequence of the phase shift is markedly different in the laser cavity compared to free propagation. The cavity responds by resonating at a different frequency, thereby amplifying a different part of the gain medium's emission line and ultimately shifting the laser frequency. Notably, this frequency shift does not scale with a positive power of $2\pi f_g \ell_c$ when $2\pi f_g  \ell_c \ll 1$ but rather approaches a constant, $-\frac{1}{2} h_+ s_\vartheta^2 \cos[\varphi_0(t)]$. During free propagation, on the other hand, photons gradually build up a frequency shift, therefore \cref{eq:fshift_freefall} scales as $\frac{1}{2} h_+ (2 \pi f_g L) s_\vartheta^2 \sin[\varphi_0(t)]$ in the limit $2\pi f_g L \ll 1$.

We finally remark that the frequency of the cesium absorption line is not significantly affected by the GW because a correction to the gravitational force can be safely ignored compared to electromagnetic forces when computing atomic transition energies \cite{Siparov}.

%=============================================================================
\section{Free-Fall Condition}
\label{sec:free-fall}
%=============================================================================

\Cref{eq:fshift-laser,eq:fshift_freefall} are based on the assumption that the components of the optical setup are free-falling, that is, they respond to the effective mechanical force exerted by the gravitational wave without counteracting it. This is a good approximation at sufficiently high frequencies, even when the apparatus is firmly affixed to an optical table. The reason is that the characteristic mechanical frequencies of any realistic support system are then well below $f_g$, which implies that the mount can be viewed as a harmonic oscillator driven at a frequency far above its resonance frequencies. Its response to the high-frequency driving force is then the same as for a free-floating object~\cite{Bringmann:2023gba}. In this context stress that the frequency shift $\Delta f_{\gamma, \text{laser}}$ is \textit{not} due to mechanical excitation of the cavity, but exclusively due to impact of the GW on photon propagation in the laser cavity.

More precisely, the authors of Ref.\,\cite{Gue:2026kga} have shown that, for a roughly cubic object of size $\ell$, the condition for free-fall is
\begin{align}
    2\pi f_g \ell \gg v_s (Q_{m0})^{1/3} \,,
    \label{eq:ff-condition}
\end{align}
where $v_s$ is the speed of sound and $Q_{m0}$ is the quality factor (i.e., the sharpness) of the object's fundamental mechanical resonance. If \cref{eq:ff-condition} is satisfied, the displacement field $\delta \vec{x}$ of the object is strongly suppressed in the TT frame. For typical values, $\ell \simeq \SI{10}{cm}$, $v_s = \SI{5000}{m/s}$, $Q_{m0} \simeq 100$, free-fall is a good approximation at frequencies above $\text{few} \times \SI{10}{kHz}$. We will need to take this into consideration when analyzing actual experimental data below.

%=============================================================================
\section{Experimental Results}
\label{sec:exp}
%=============================================================================

In the following, we set new experimental limits on HFGW using data first presented in Ref.\,\cite{Tretiak:2022ndx} in the context of a search for ultralight dark matter. Two different experimental setups were employed, both following the general layout depicted in \cref{fig:exp-sketch}. In both setups, cesium vapor cells were illuminated by a Ti:Sapphire ring laser, with the laser frequency tuned to the slope of a particular hyperfine component of the $6^2S_{1/2} \to 6^2P_{3/2}$ atomic transition (\SI{351.72196}{THz}, natural line width $\sigma_\gamma \simeq \SI{5.5}{MHz}$). Due to the sharpness of the atomic lines, a small change in the frequency difference between the incoming photons and the atomic resonance has a large impact on the transmissivity, which is the measured observable. One setup (``Experiment~A'') employed Doppler-free polarization spectroscopy, while the other (``Experiment~B'') used a Doppler-broadened line (full-width at half maximum $\simeq \SI{420}{MHz})$. This way, Experiment~A offered higher sensitivity to small frequency shifts, while Experiment~B offered a larger bandwidth. A detailed description of the experimental setups can be found in the supplemental material of Ref.\,\cite{Tretiak:2022ndx}.

\begin{figure*}
    \centering
    \includegraphics[width=0.7\textwidth]{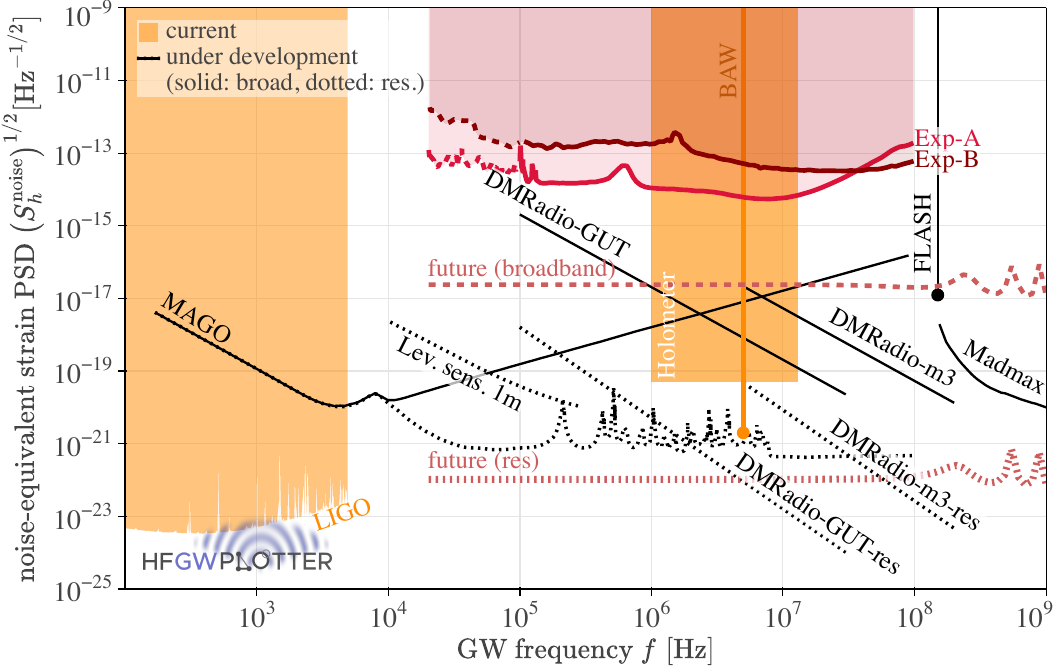}
    \caption{Constraints on high-frequency gravitational waves from precision spectroscopy of cesium atoms \cite{Tretiak:2022ndx}. Red solid: current constraints from Experiments~A and B discussed in this paper, assuming optimal GW direction and polarization (see text for details); below \SI{100}{kHz}, the curves are dashed to emphasize that in this regime the free-fall condition from \cref{eq:ff-condition}, on which our analysis is based, is no longer strictly satisfied. Red dotted: expected future improvements in broadband mode (top), and in resonant mode with an $L$-shaped geometry (bottom). We compare to other experimental limits (orange shaded) and to sensitivities of experiments under active development (black) as collected in Ref.~\cite{Aggarwal:2025noe}. Plot created using a modified version of HFGWPlotter \cite{Muia:2025}.
    }
    \label{fig:h-vs-f}
\end{figure*}

Using \cref{eq:fshift-laser,eq:fshift_freefall}, it is straightforward to convert the constraints on frequency shifts $\Delta f / f$ obtained in Ref.\,\cite{Tretiak:2022ndx} into limits on the amplitude of a hypothetical high-frequency gravitational wave signal. The result as a function of GW frequency is shown in red in \cref{fig:h-vs-f} for Experiments~A and B, as well as for hypothetical future setups discussed below. Over most of the frequency range, the GW effect on the laser cavity (\cref{eq:fshift-laser}) dominates the limit; only at $f_g \gtrsim \SI{10}{MHz}$, the frequency shift during propagation (\cref{eq:fshift_freefall}) becomes important as well. Note that at frequencies below \SI{100}{kHz}, the free-fall condition in \cref{eq:ff-condition} is no longer fully satisfied, so our limit should only be considered indicative in this regime. A full calculation taking into account deviations from free-fall could lead to either a weaker limit (as the mechanical forces in the setup counteract the GW force) or to a stronger limit (as mechanical resonances could be excited).

In \cref{fig:h-vs-f} we show constraints in terms of the strain-equivalent noise power spectral density (PSD),
\begin{align}
    (S_h^{\rm noise})^{1/2} = h^{\rm max,coh} \sqrt{t_{\rm obs}} \,.
\end{align}
Here $h^{\rm max,coh}$ is the maximum allowed GW strain for a signal that, closely mimicking the ultralight dark matter case, is coherent over the full observation time $t_{\rm obs}$ -- which most sources of high-frequency GWs are not. This is the reason we choose to plot $(S_h^{\rm noise})^{1/2}$ instead: $(S_h^{\rm noise})^{1/2}$ is a property of the detector that is independent of any assumptions on the signal.
In particular, $S_h^{\rm noise}$ can be understood as the observed experimental noise, folded with the inverse of the detector response function.

There is an important caveat related to the way the data in Ref.~\cite{Tretiak:2022ndx} were analyzed: peaks that were not visible in both Experiment~A and Experiment~B were removed from the data. As the two experiments were not run simultaneously, this would have eliminated any transient signal (primordial black hole merger, supernova, etc.). Therefore, the limits displayed in \cref{fig:h-vs-f} apply as such only to persistent GW signals, whereas they should be interpreted as the possible reach of this setup for transient GW signals if the two experiments were run simultaneously.

This motivates a closer look at persistent GW signals, like superradiant boson clouds around black holes or primordial black hole binaries in their inspiral stage. For black hole superradiance, GW emission in the MHz to GHz band would require new ultralight bosons in the $10^{-9}$--$\SI{e-6}{eV}$ mass range and primordial black holes with masses $\sim 10^{-4}$--$10^{-1} M_\odot$. For such persistent sources, the signal-to-noise ratio can be accumulated over many GW oscillation periods. We can thus directly translate the limits from ultralight dark matter searches to bounds on $h^\text{max,coh}$, in particular
\begin{align}
    h^\text{max,coh} \lesssim \num{e-17}
    \qquad
    \text{($f_g \sim \SI{100}{kHz}$ to \SI{10}{MHz})} \,.
    \label{eq:bound-coherent}
\end{align}
In principle, our setup is also sensitive to processes in the early Universe which produce stochastic GW backgrounds. However, we remind the reader that any such source strong enough to saturate our current limits would violate constraints on the total relativistic energy density in the Universe by many orders of magnitude~\cite{Cyburt:2015mya, Yeh:2022heq, Planck:2018vyg, Goldstein:2026iuu}.

Like for any GW detector, the sensitivity of our experiments depends on the arrival direction of the gravitational wave relative to the orientation of the apparatus. In \cref{fig:h-vs-f} and \cref{eq:bound-coherent}, we have assumed an optimal configuration, where the GW travels orthogonal to the photon trajectory ($\vartheta = \pi/2$). For general arrival direction $\vartheta$, our limits would be rescaled by a factor
\begin{align}
    s \simeq 1 - c_\vartheta \frac{\sin[ 2\pi f_g \ell_c c_\vartheta ]}{\sin[ 2\pi f_g \ell_c ]}
    \label{eq:theta-rescaling}
\end{align}

As evident from \cref{fig:h-vs-f}, our current limits cannot compete with dedicated GW searches (namely the Fermilab holometer \cite{Holometer:2016qoh} and the bulk acoustic wave device described in Ref.\,\cite{Goryachev:2021zzn}) in the frequency range which they cover. However, the sensitivity of our experiments covers a much broader frequency range. This is particularly important given the large parameter space to explore for high frequency gravitational waves and the expected nature of many of these signals, which could be burst-like or sweep through frequency space. The  results presented here constitute the first experimental constraints on gravitational waves between \SI{13}{MHz} and \SI{100}{MHz}.

%=============================================================================
\section{Future Sensitivity}
\label{sec:sens}
%=============================================================================

While the limits shown in \cref{fig:h-vs-f} already probe previously unconstrained regions of HFGW parameter space, the techniques employed there leave ample room for improvement. In particular, the current experimental noise levels are still above the fundamental limits, but there are paths forward to reach these limits, as we discuss below (see also Refs.~\cite{Tretiak:2022ndx, Heisig:2025oim}).

To estimate the sensitivity achievable in the future, we distinguish measurements that are sensitive to the GW amplitude ($\propto h$) and those sensitive to the GW power ($\propto h^2$). We call these \emph{linear} and \emph{quadratic} measurements, respectively.
Spectroscopic experiments like the ones discussed here can fall into either category:

\textbf{Linear measurements.}
If the frequency of the laser beam in the setup from \cref{fig:exp-sketch} is tuned to the rising or falling edge of an atomic absorption line, a small shift $\Delta f_\gamma$ in the photon frequency leads to a proportional change in the measured transmitted light power, $\Delta P \propto \Delta f_\gamma / \sigma_\gamma$, where $\sigma_\gamma$ is the linewidth. If the sampling frequency, $f_s$, is higher than twice the GW frequency, $f_g$, the GW manifests as a time-dependent modulation of the transmitted power, and therefore the measured photon frequency. The amplitude of this modulation is $\Delta f_\gamma / f_\gamma \sim h$ (see \cref{eq:fshift-laser,eq:fshift_freefall}), hence the observable signal is linear in the strain. This shows that the experiments described in the previous section fall into the linear category.

\textbf{Quadratic measurements.} If, in contrast, $f_s < 2f_g$, the GW effect instead manifests as sidebands in the photon spectrum carrying a fraction $\epsilon_{\rm sb} \simeq \frac{1}{4}(h\,f_\gamma/f_g)^2$ of the power \cite{Bringmann:2023gba}, quadratic in strain. In the following, we focus on linear measurements only.

\vspace{1em}
We now estimate the degree of improvement that is possible in an experiment similar to the one in \cref{fig:exp-sketch} by reaching the standard quantum limit, with shot noise as the dominant noise source. The total number of photons collected over an observation time $t_{\rm obs}$ is
\begin{align}
    N_\gamma = \frac{P \, t_{\rm obs}}{2\pi f_\gamma} 
               \approx 10^{23} 
               \bigg( \frac{P}{\SI{1}{W}} \bigg) 
               \bigg( \frac{t_{\rm obs}}{\SI{10}{hr}} \bigg) \, .
\end{align}
If the signal is persistent, as is the case for instance for superradiance or cosmological sources, $t_{\rm obs}$ is simply the total data taking time. For transient signals such as primordial black hole mergers or supernovae, it is instead understood to be the -- typically much shorter -- duration of the signal event. The photon shot-noise-limited sensitivity of a single transmitted power measurement sample can be obtained by setting $\Delta P/P_0 = 1/\sqrt{N_s}$, where $P_0$ is the input light power, $N_s = P / (2\pi f_\gamma f_s)$ is the number of detected photons per sample. This leads to a single-sample sensitivity $h \gtrsim (\sigma_\gamma / [2\pi f_\gamma]) \sqrt{2\pi f_s f_\gamma / P}$. For a GW signal at frequency $f_g < f_s$, the $N_{\rm samp} = f_s t_{\rm obs}$ independent samples can be combined, improving the sensitivity by $\sqrt{N_{\rm samp}}$ \cite{Aggarwal:2025noe}.  The minimum detectable strain is therefore of order\footnote{We assume here that the number of irradiated atoms is sufficiently large for noise due to fluctuations in their number to be negligible.}
\begin{align}
    &h^{\rm linear}
        \sim \frac{\Delta f_\gamma}{f_\gamma} \Big|^{\rm linear}
        \gtrsim \frac{\sigma_\gamma}{2\pi f_\gamma}
                 \sqrt{\frac{2\pi f_\gamma}{P \, t_{\rm obs}}} \notag\\
        &\!\!\approx \num{6.8e-19} \times\!
                 \bigg(\! \frac{\sigma_\gamma}{\si{GHz}} \bigg) \!
                 \bigg(\! \frac{\SI{e15}{Hz}}{f_\gamma} \bigg)^{\!\!1/2} \!\!
                 \bigg(\! \frac{\SI{1}{W}}{P} \bigg)^{\!\!1/2} \!
                 \bigg(\! \frac{\SI{10}{hr}}{t_{\rm obs}} \bigg)^{\!\!1/2} \!\!,
    \label{eq:linear-sens}
\end{align}
\Cref{eq:linear-sens} holds only if $f_s > f_g$ as only in this case the time-evolution of the GW signal can be resolved. For $f_s < f_g$, the signal will manifest in the form of sidebands.

\vspace{1em}
Assuming a $\Delta f_\gamma / f_\gamma$ sensitivity given by \cref{eq:linear-sens} with the parameters and geometry of Experiment~A from Ref.\,\cite{Tretiak:2022ndx} ($P = \SI{10}{mW}$, $f_\gamma = \SI{3.5e14}{Hz}$, $\sigma_\gamma = \SI{5.5}{MHz}$), we obtain the red dashed sensitivity projection labeled ``future (broadband)'' in \cref{fig:h-vs-f}.

A further sensitivity increase is achievable by replacing the Cs vapor cell with a high-finesse optical cavity. To avoid averaging the frequency shift to zero as photons dwell in the cavity for many GW periods, the cavity will need to have a non-trivial geometry tuned to the GW frequency. For instance, an L-shaped cavity can shield photons from the GW during part of each round-trip, causing a net phase shift that translates into a shift of the cavity's resonance frequency according to \cref{eq:fshift-laser}. This allows the experiment to benefit from the cavity's narrow linewidth even at high GW frequencies. For such a setup with $\sigma_\gamma = \SI{1}{kHz}$ and $P = \SI{10}{W}$, \cref{eq:linear-sens} yields the ``future (res)'' curve in \cref{fig:h-vs-f}.

Finally, given the relative simplicity of the proposed setups, it is conceivable to replicate them many times. Such a detector network would provide an additional handle for reducing systematic uncertainties, it would improve angular coverage if different detectors are oriented differently, and it will directly improve the sensitivity, typically by a factor $\propto \sqrt{N}$, where $N$ is the number of detectors \cite{Aggarwal:2025noe, Amaral:2026bef}.

Further details on future improvements of the experimental setups used in this paper are given in the Appendix.

%=============================================================================
\section{Discussion and Outlook}
\label{sec:conclusions}
%=============================================================================

In summary, we have discussed spectroscopic searches for high-frequency gravitational waves in the MHz--GHz range. These searches exploit the frequency modulation which photons experience when travelling through a GW background, both within and beyond the laser cavity. Due to the boundary conditions of the laser cavity, the former is significantly more sensitive in most of the frequency range considered.
We have then used data from a spectroscopic search for ultralight dark matter (Ref.~\cite{Tretiak:2022ndx}) to derive new limits on persistent gravitational wave backgrounds from 0.1--\SI{100}{MHz}, see \cref{fig:h-vs-f}. Over much of this frequency range, these are the first experimental constraints on high-frequency gravitational waves. While no realistic sources are expected to have amplitudes large enough to saturate our current limits, the results presented here constitute an important proof of principle. We have outlined how we plan to improve the sensitivity by up to eight orders of magnitude to reach a strain-equivalent noise level
\begin{align}
    (S_h^{\rm future})^{1/2} \sim \SI{e-22}{Hz^{-1/2}} \,.
\end{align}

%=============================================================================
\section*{Acknowledgments}
%=============================================================================

This work has been supported by the Cluster of Excellence ``Precision Physics, Fundamental Interactions, and Structure of Matter'' (PRISMA++ EXC 2118/2) funded by the German Research Foundation (DFG) within the German Excellence Strategy (Project ID 390831469), by the COST Action within the project COSMIC WISPers (Grant No.\ CA21106), and by ERC grant ERC-2024-SYG 101167211 grant (GravNet).

%=============================================================================
\appendix
\section{Improved Spectroscopic High-Frequency Gravitational Wave Searches}
\label{sec:future}
%=============================================================================

In the following, we discuss in more detail the possible future improvements of the experimental setups discussed in the main part of the paper.

\begin{figure*}
    \begin{tabular}{@{}cc@{}}
      \includegraphics[width=0.48\textwidth]{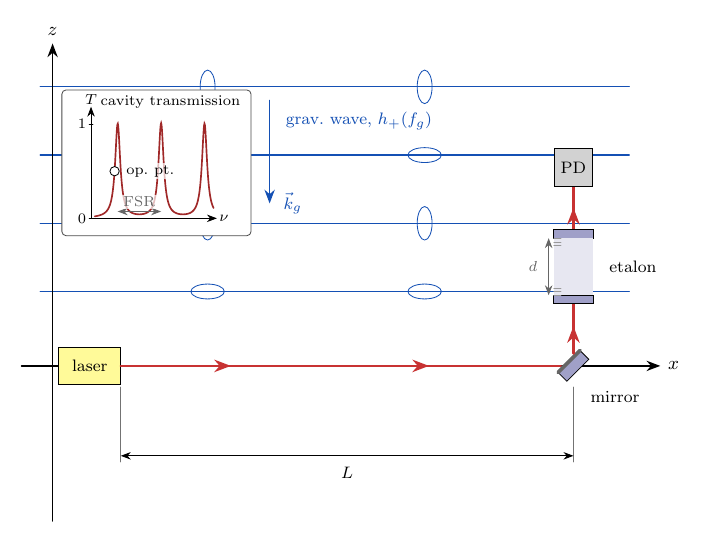} &
      \includegraphics[width=0.48\textwidth]{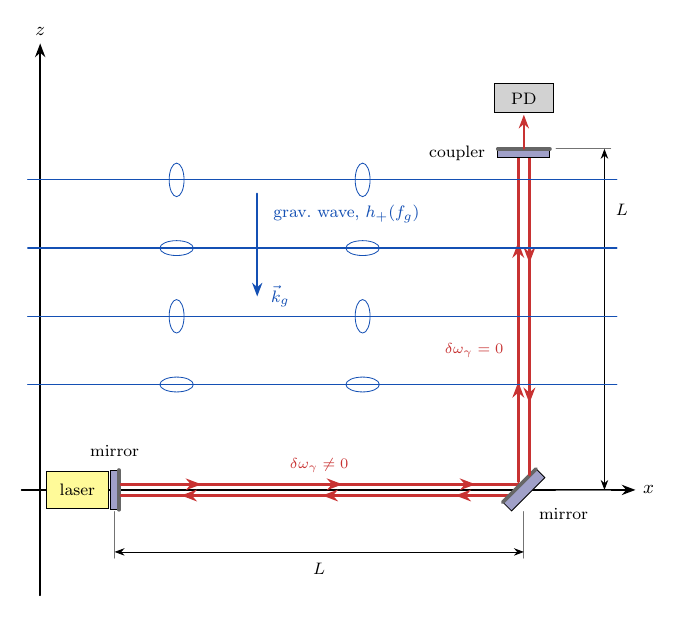} \\
      \textbf{(a)} & \textbf{(b)} \\
      \includegraphics[width=0.48\textwidth]{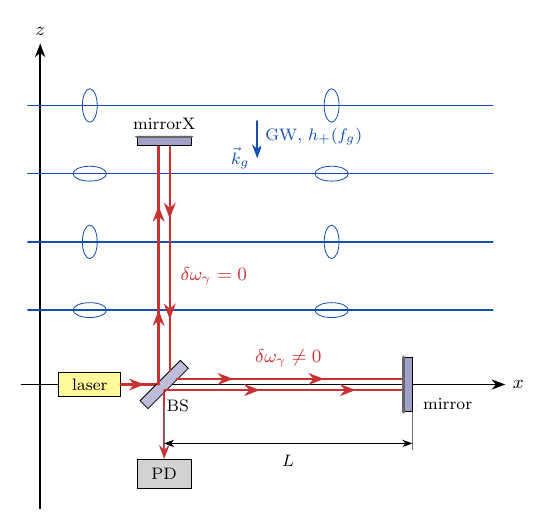} &
      \includegraphics[width=0.48\textwidth]{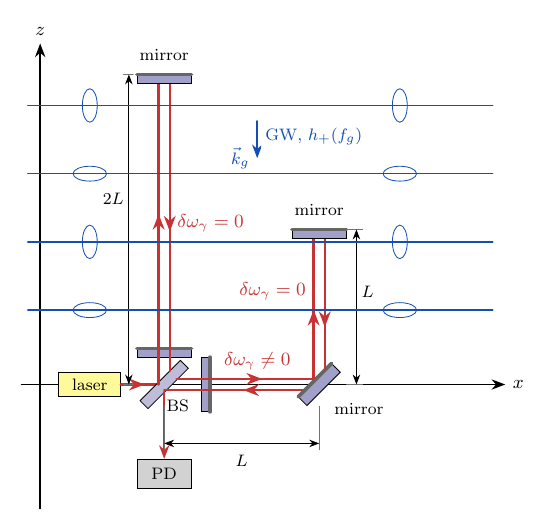} \\
      \textbf{(c)} & \textbf{(d)}
    \end{tabular}
    \caption{Proposed experimental configurations for precision spectroscopic GW searches. All panels show a laser beam (red) and a gravitational wave $h_+(f_g)$ propagating along $z$ (horizontal blue wavefronts with \smash{$\vec{k}_g$} arrow); a photodetector (PD) records the output signal.
    \textbf{(a)}~Basic etalon-based setup: the laser propagates along a baseline $L$ before passing through a Fabry--P\'{e}rot etalon acting as a narrow-band frequency filter.
    \textbf{(b)}~More advanced etalon-based setup with an L-shaped optical cavity: photons accumulate a phase shift in the horizontal arm ($\Delta f_\gamma \neq 0$) while remaining unaffected in the vertical arm ($\Delta f_\gamma = 0$). This leads to a net shift in the cavity's resonance frequency.
    \textbf{(c)}~Standard Michelson interferometer: arm~1 (along $z$) acquires no frequency shift ($\Delta f_\gamma = 0$), while arm~2 (along $x$) acquires a shift ($\Delta f_\gamma \neq 0$); the differential phase is recorded at the dark port. 
    \textbf{(d)}~Michelson interferometer with an L-shaped arm: arm~1 is straight (along $z$, $\Delta f_\gamma = 0$) and arm~2 is L-shaped, with a horizontal section acquiring a frequency shift and a vertical section that does not, enabling the same resonant enhancement as in panel~(a).
    }
    \label{fig:future-setups}
\end{figure*}

%-----------------------------------------------------------------------------
\subsection{A Different Optical Resonator}
\label{sec:etalon}
%-----------------------------------------------------------------------------

To eliminate noise due to fluctuations in the number of irradiated atoms in the cesium vapor cell, the latter can be replaced with a different type of frequency filter, namely by an etalon (Fabry--P\'erot cavity), see \cref{fig:future-setups}\,(a). In the same way that, for the vapor cell, the laser frequency has been tuned to the rising (or falling) edge of an atomic transition, here it would be tuned to the rising (or falling) edge of a cavity mode, again maximizing the change in transmissivity caused by small, GW-induced photon frequency variations.

An important consideration in this case is that, like for a laser cavity, also the resonance frequency of an etalon will be shifted by a passing GW. In this sense, the etalon-based setup is conceptually different from the one involving a cesium vapor cell, whose resonance frequency remains unaffected by the GW. The GW effect on the etalon needs to be taken into account when interpreting the measured transmissivity modulation.

For a high-finesse etalon, the effective photon retention time in the cavity is larger than the GW period. In this case, the shift of the etalon resonance is suppressed, and the GW effect on the etalon can be neglected. (The resonance does develop sidebands in this case, but they do not affect our measurement.)

%-----------------------------------------------------------------------------
\subsection{Integrating over Multiple GW Periods}
\label{sec:L-shaped}
%-----------------------------------------------------------------------------

\Cref{fig:future-setups}\,(b) shows a sketch of the setup briefly described below \cref{eq:linear-sens} in the main text (see also Ref.\,\cite{Heisig:2025oim} for related ideas). It extends the ideas from \cref{sec:etalon} by giving the frequency-analyzing etalon an L-shaped geometry. The optical cavity is formed by three mirrors (two to delimit the cavity, one to deflect the photon at the midpoint of the cavity). Assume the arm length is tuned to the gravitational wave length $\lambda_g$ according to $L = \lambda_g / 4$, and assume that a photon enters the cavity when the GW phase is $\varphi_0$. Assume moreover that the GW travels along the $z$ direction. Following Ref.~\cite{Domcke:2024abs} one can show in analogy to the derivation of \cref{eq:fshift-laser,eq:fshift_freefall} that the photon accumulates a phase shift $\Delta \phi = h_+ f_\gamma / (2 f_g) \times \big[ \sin(\varphi_0) - \cos(\varphi_0) \big]$ while propagating along the cavity's horizontal arm in the positive $x$-direction  ($\vartheta = \pi/2$). When the GW phase has reached $\varphi_0 + \pi/2$, the photon enters the vertical arm, where the angle between the GW and photon directions is $\vartheta = \pi$. The photon does not accumulate any further phase shift during this quarter-period. The same applies to its way back through the vertical arm ($\vartheta=0$). But when the photon is back in the horizontal arm ($\vartheta = -\pi/2$, GW phase $\varphi_0 + \frac{3}{2} \pi$), its phase is again shifted, this time by $\Delta \phi = h_+ f_\gamma / (2 f_g) \times \big[ -\sin(\varphi_0) - \cos(\varphi_0) \big]$. The cycle then repeats.

As a consequence, the cavity's resonance frequency is shifted by\footnote{We obtain this formula by using Eq.\,(3.9) from Ref.\,\cite{Domcke:2024abs} for the phase shift in a single segment of the cavity, and then adding up the phase shifts from the four segments that make up one round-trip in an L-shaped cavity. We then use $\Delta f_\gamma / f_\gamma = \Delta\phi / (4 L \times 2\pi f_\gamma)$.}
\begin{align}
    \frac{\Delta f_\gamma}{f_\gamma} \Big|_{\rm cavity}
      = -\frac{h_+}{2\pi} \cos(\varphi_0) \,.
    \label{eq:fshift-L-geometry}
\end{align}
Crucially, thanks to the special geometry, photons experience this shift even though each individual photon is retained in the cavity for many GW cycles. This is possible since the phase shift depends on the angle between the photon and GW propagation direction, and hence can `hide' from the GW effect in one of the arms. The case discussed above is the optimal case, in which one arm shows the maximal response while the other features zero response. For a general direction of the incoming GW the effect is suppressed, though vanishes only for specific GW directions. This allows the experiment to benefit from a high-finesse cavity even at high GW frequency. For instance, a resonance line width of order kHz should be achievable, which would lead to substantial sensitivity improvements, see \cref{fig:h-vs-f} above.

%-----------------------------------------------------------------------------
\subsection{Detector Networks}
\label{sec:networks}
%-----------------------------------------------------------------------------

Given the simplicity of the proposed setups, it is conceivable to replicate them many times. First, such a detector network would provide an additional handle for reducing systematic uncertainties. In fact, this was already exploited in the experiments discussed in \cref{sec:exp} above, where peaks that were seen only in Experiment~A or only in Experiment~B were removed. Second, we already discussed in the context of \cref{fig:future-setups}(a,b) the benefits of operating several detectors oriented differently to optimize angular coverage. Finally, multiple detectors oriented the same way will directly improve the sensitivity: for a network of $N$ detectors, the strain-equivalent noise scales as $\sqrt{S_h} \propto 1/\sqrt{N}$ \cite{Aggarwal:2025noe}. Should it be possible to set up a network of quantum-entangled sensors, the scaling would improve to $\sqrt{S_h} \propto 1/N$, see for example \cite{Amaral:2026bef} and references therein.

Beyond simple replication of identical setups, substantial sensitivity gains may be achieved through structured arrays of sensing elements arranged in linear, planar, or even three-dimensional geometries (for example, a cubic network). In such architectures, multiple channels would sample the gravitational-wave induced phase shift at different spatial locations. Coherent comparison of these channels could enhance the common signal, suppress uncorrelated technical noise, and provide directional information on the incoming wave through relative phase delays. Modern multi-channel digital readout and correlation techniques developed for phased-array radar, radio astronomy, and optical metrology suggest that scaling to large numbers of channels is technically feasible.

%-----------------------------------------------------------------------------
\subsection{Interferometry}
\label{sec:interferometry}
%-----------------------------------------------------------------------------

The GW-induced photon frequency shifts studied in this paper represent the same physics as the phase shifts observed in interferometric gravitational wave detectors like LIGO / VIRGO / KAGRA, LISA and the Fermilab Holometer \cite{Domcke:2024eti}. It is therefore reasonable to explore the prospects of advanced interferometric techniques at kHz--GHz frequencies. The basic layout of the detector would remain the same as in lower-frequency GW searches, but scaled down to a size comparable to the GW wavelength. The simplest option is a Michelson interferometer, as sketched in \cref{fig:future-setups}(c).

Assuming a shot noise-limited measurement, we can estimate the sensitivity of such a setup as follows: a GW with strain $h$ will change the arm length by $h \, L$, and therefore the photon phase by $2\pi h \, L \, f_\gamma$. Assuming a working point on the rising or falling edge of an interference fringe, phase shifts of order $1/\sqrt{N_\gamma}$ are observable, where $N_\gamma$ is the total number of observed photons. It can be expressed in terms of the laser power, $P$, and the observation time $t_{\rm obs}$ as $N_\gamma = P \, t_{\rm obs} / (2\pi f_\gamma)$. The sensitivity of an interferometer is therefore of order
\begin{align}
    h &\gtrsim \frac{1}{L \sqrt{2\pi P \, t_{\rm obs} \, f_\gamma}}
                                \notag\\
      &= 10^{-19} \times
         \bigg( \frac{\SI{30}{cm}}{L} \bigg)
         \bigg( \frac{\SI{1}{W}}{P} \bigg)^{1/2}
         \bigg( \frac{\SI{10}{hr}}{t_{\rm obs}} \bigg)^{1/2} \,.
    \label{eq:sens-interferometer}
\end{align}

Further sensitivity improvements are possible (at the expense of reduced bandwidth) by turning the interferometer arms into optical cavities, as in LIGO, VIRGO and KAGRA, but with a geometry following the same principles as the one discussed above in \cref{sec:L-shaped}. An example is shown in \cref{fig:future-setups}(d): one interferometer arm is formed by a cavity with the same L-shaped geometry as in \cref{fig:future-setups}(a), while the other is straight. For a GW traveling along the $z$ direction, photons in the straight arm will remain unperturbed, while those in the L-shaped arm experience a frequency shift, or, equivalently, a phase shift that affects the interference pattern at the output port and therefore the light intensity registered by the photon detector.

%=============================================================================
\subsection{Practical Considerations}
\label{sec:compact-fp-implementation}
%=============================================================================

\begin{figure}[t]
    \centering
    \includegraphics[width=\linewidth]{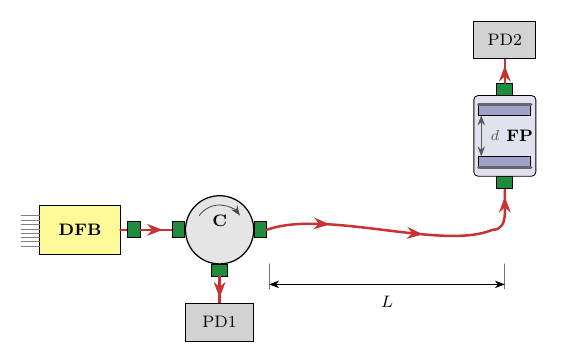}
    \caption{Conceptual compact implementation of a HFGW sensor based on a Fabry--P\'erot cavity. Light from a DFB laser is sent through an optical circulator into a compact fiber Fabry--P\'erot resonator. The reflected signal is separated by the circulator and detected on one photodiode, while the transmitted signal is monitored on the other. The laser is locked to an operating point on the slope of the Fabry--P\'erot resonance where the reflected and transmitted powers are approximately equal. HFGW-induced frequency modulation of the laser output and of the cavity resonance frequency is converted into a differential intensity signal.
    }
    \label{fig:compact-fp-detector}
\end{figure}

Building on the Fabry--P\'erot (FP) frequency-discriminator scheme discussed above, see \cref{fig:future-setups}(b), we can envision a simple and scalable implementation based entirely on standard commercial fiber-optic telecommunication components. A possible layout is shown in \cref{fig:compact-fp-detector}. The light source is a compact distributed-feedback (DFB) diode laser operating at a standard telecommunication wavelength, such as \SI{1310}{nm} or \SI{1550}{nm}. Such lasers are commercially available in fiber-coupled butterfly packages and can be directly connected to standard single-mode fiber components. The laser output is sent to an optical circulator, which routes the light to a compact fiber FP resonator and directs the reflected light to one port of a balanced photodetector. The transmitted light is monitored on the second port. Fiber-optic setups of this type have been deployed in analogous ultralight dark matter searches~\cite{Manley:2023usr, Savalle:2020vgz}.

The FP resonator is operated on the slope of one of its resonances, at a point where the reflected and transmitted powers are approximately equal,
\begin{align}
    P_T \simeq P_R .
\end{align}
For a low-loss FP resonator, a small detuning from resonance produces changes in transmission and reflection with approximately opposite signs,
\begin{align}
    \delta P_T \simeq -\delta P_R .
\end{align}
The balanced difference signal is therefore close to zero at the operating point, but has first-order sensitivity to frequency fluctuations. The sum signal monitors the total optical power,
\begin{align}
    P_{\mathrm{diff}} = P_T - P_R\,,
    \qquad
    P_{\mathrm{sum}} = P_T + P_R\,.
\end{align}
In practice, the readout can be implemented using a balanced photodetector, which can enhance the frequency-discriminator signal while suppressing common-mode laser-intensity noise and compensating, to first order, unequal gains in the two photodetection channels.

The same differential signal can be used as the error signal for locking the DFB laser to the FP resonator. The feedback loop should be slow compared to the gravitational-wave search band. For example, a servo bandwidth well below \SI{20}{kHz} would stabilize slow drifts and suppress acoustic and mechanical disturbances, while not affecting MHz-frequency and higher gravitational-wave-induced modulation in the balanced-detector output. Thus the lock maintains the operating point without tracking the signal frequencies of interest.

The main practical advantage of this architecture is its scalability. The required components, including telecom DFB lasers, circulators, fiber FP resonators, InGaAs photodiodes, balanced receivers, and high-speed digitizers, are compact, commercially available, and benefit from large-scale industrial production. Some selection of laser diodes may be required to obtain sufficiently low frequency noise and good long-term stability. If many channels are built, such component selection can be performed at the array level while keeping the average cost per channel moderate.

This makes it conceivable to form a local array with tens to hundreds of compact FP sensors at moderate cost. Such a local array would be distinct from a geographically distributed detector network: all sensors could be operated in the same laboratory or facility, sharing common infrastructure for synchronization, environmental monitoring, and data acquisition. Correlating the outputs of multiple sensors would suppress uncorrelated technical and electronic noise and improve the strain sensitivity of the array. In addition, FP resonators oriented along different spatial directions would sample different projections of the gravitational-wave perturbation. A multi-orientation local array could therefore improve angular coverage, reduce blind spots of a single-detector geometry, and provide information about the propagation direction and polarization content of the gravitational wave. Further benefits may be gained by creating geographically separated ``supernetwork'' of such local network nodes, see the discussion in \cite{Amaral:2026bef}.

Note that the response of a setup involving fiber-optic connections must take into account the refractive index, $n$, of the fiber medium. In principle, a GW could affect $n$ by modulating the density of the medium. This could be relevant for media with sizable refractive index.

%-----------------------------------------------------------------------------
\bibliography{refs}
%-----------------------------------------------------------------------------

\end{document}